\documentclass[notitlepage,reprint,onecolumn, amsmath,amssymb,a4paper, aps,prd, nofootinbib]{revtex4-2} 
\usepackage{comment}
\usepackage{graphicx}
\usepackage{bm}
\usepackage{color}
\definecolor{ao}{rgb}{0,.5,.15}
\usepackage{amssymb}
\usepackage{amsmath}
\usepackage[utf8]{inputenc} 
\usepackage[T1]{fontenc}    

\newcommand{\bit}{\begin{itemize}}
\newcommand{\eit}{\end{itemize}}
\newcommand{\ben}{\begin{enumerate}}
\newcommand{\een}{\end{enumerate}}
\newcommand{\beq}{\begin{equation}}
\newcommand{\eeq}{\end{equation}}
\newcommand{\bea}{\begin{eqnarray}}
\newcommand{\eea}{\end{eqnarray}}
\newcommand{ \lsim}{\mathrel{\vcenter
     {\hbox{$<$}\nointerlineskip\hbox{$\sim$}}}}

\newcommand{\gappeq}{\mathrel{\rlap {\raise.5ex\hbox{$>$}}
{\lower.5ex\hbox{$\sim$}}}}
\newcommand{\lappeq}{\mathrel{\rlap{\raise.5ex\hbox{$<$}}
{\lower.5ex\hbox{$\sim$}}}}

\def\bq{\begin{quote}}
\def\eq{\end{quote}}

\def\Dslash{ \, D  \! \! \! \! / ~ }
\def\Aslash{ \, A  \! \! \! \! / ~ }
\def\dslash{ \, \partial  \! \! \! \! / ~ }

\def\LNP{\Lambda_{NP}}

\def\meg{\mu \to e \gamma}

\def\teg{\tau \to e \gamma}
\def\tmg{\tau \to \mu \gamma}

\def\tlg{\tau \to l \gamma}

\def\tlnn{\tau \to l \bar{\nu} \nu}
\def\menn{\mu \to e \bar{\nu} \nu}

\def\a{\alpha}
\def\b{\beta}
\def\g{\gamma}
\def\d{\delta}

\def\m{\mu}

\def\r{\rho}
\def\s{\sigma}

\begin{document}

\renewcommand{\thefootnote}{\fnsymbol{footnote}}

\begin{center}
{\Large \textbf{
Left-Handed Physics is not right for EDMs}}

\vskip 20pt
{\large Marco Ardu$^1$\footnote{E-mail address: marco.ardu@ific.uv.es },
  Sacha Davidson$^2$\footnote{E-mail address: s.davidson@ip2i.in2p3.fr},
  and Nicola Valori$^1$\footnote{E-mail address: nicola.valori@uv.es}
}
	
	\vskip 10pt  
{\it $^1$
    Instituto de F\'isica Corpuscular, Universidad de Valencia and CSIC, Edificio Institutos Investigaci\'on, C/Catedr\'atico
Jos\'e Beltr\'an 2, 46980 Paterna, Spain}\\	
{\it $^2$Universite Claude Bernard Lyon 1, CNRS/IN2P3, IP2I Lyon, UMR 5822, Villeurbanne, F-69100, France}\\

	\vskip 20pt

{\small \bf Abstract}\\
\end{center}
{\small
Heavy New Physics models  with lepton flavour-changing interactions are motivated by neutrino masses, and generically induce dipole interactions for leptons---which can be flavour-changing ($l_j\to l_i \gamma$), or flavour-diagonal (magnetic and electric dipole moments(edms)).
We focus on models with complex couplings, and where
the singlet  Standard Model leptons  ($\{e_R^i\}$)
do not interact with the New Physics.
In such models, edms  are calculated to arise at two loops,  despite that complex amplitudes for $l_j\to l_i \gamma$ appear at one loop.
We explore whether the extra loop suppression of edms  survives flavour basis rotations that could be induced by  flavour-changing NP contributions to the charged lepton mass matrix.
We show that  one-loop edms vanish in  both the mass and Yukawa eigenstate bases.

}


$~$\\


\section{Introduction}
\label{sec:intro}

 Electric Dipole Moments (edms) \cite{Dar:2000tn,Pospelov:2005pr,Safronova:2017xyt,Vives:2025clr}  have yet to be observed,  so the  experimental searches for these CP-violating observables
probe very heavy new particles  with complex couplings constants.
The Standard Model (SM) expectations for  edms \cite{Pospelov:1991zt,Yamaguchi:2020dsy,Yamaguchi:2020eub} are far below the current experimental sensitivity, so  an observation in upcoming searches would be an unambiguous signature of New Physics (NP).
This is especially interesting  in the case of leptonic edms, because the observed neutrino mass matrix implies that there is NP in the leptonic sector.
In the following,  this NP is assumed to be  heavy, so  that it can be parametrised with Effective Field Theory (EFT) \cite{Georgi:1993mps,Buras:1998raa,Manohar:2018aog}.  It should also  generate $[m_\nu]$ (so  it is flavour-changing), and   involve complex couplings (so  it can contribute to edms). In addition, we restrict to models  where  the new heavy mass eigenstates  do not interact with  the singlet charged leptons of the SM; some neutrino mass models satisfy this condition.

 In such models, one-loop contributions to edms appear to vanish,  and   two-loop  contributions have been   calculated with interest and care \cite{Ng:1995cs,Archambault:2004td,Abe:2013qla,Abada:2018zra,Abada:2024hpb,Fujiwara:2021vam,Fujiwara:2020unw}.  From an EFT  perspective, these model calculations usually  give a matching contribution to the diagonal dipole coefficients
\beq
d_{i} 
\sim \frac{1}{(16 \pi^2)^2} \frac{ m_i}{\LNP^2}\Im m\{ \Pi_{ii}\}
\label{estimate}
\eeq
where $\Im m\{\Pi_{ii}\}$ is the imaginary part of a product of NP  and SM couplings,  and $\LNP$ is the mass scale of the NP.

One can nonetheless wonder whether there could be other  contributions, for instance due to mass-basis redefinitions induced by NP.
The aim of this manuscript is to explore whether NP contributions  to the charged lepton mass matrix  can induce relevant contributions to edms.  The doubt arises, because the dipole operator
\bea
\d {\cal L} &=& \sum_{i,j}  \frac{ C^{ij}_{D}}{\LNP^2}
 \overline{\ell}_i H  \sigma^{\a\b} F_{\a\b} e_j + h.c
 \label{OD}
\eea
has flavour-changing off-diagonal elements which mediate  $l_j\to l_i \g$,
and  which  generically  arise at one-loop.
The  real and imaginary  parts of the diagonal elements are the  magnetic and electric dipole moments.
Since the  off-diagonals are naturally complex(for complex NP couplings), one can wonder whether a rotation to the physical mass eigenbasis could mix a one-loop  off-diagonal element into a relevant contribution to an edm.
Or whether the SM Higgs could have complex Yukawa couplings in the  lepton mass eigenstate basis, allowing it to contribute to edms.
Such contributions would be mixing-angle suppressed, but potentially comparable to two-loop diagrams.

After a brief review of dipoles in Section \ref{sec:rev}, Section \ref{sec:arg} argues  in two steps that one-loop edms vanish.
Subsection \ref{ssec:diagrams} shows  that  one-loop heavy particle  diagrams  vanish, in matching the model onto EFT at $\LNP$ in the lepton mass eigenstate basis.
Then in Subsection  \ref{ssec:RGEs}, we calculate the NP contributions to the mass matrix, {and show that  the resulting misalignment between the mass and Yukawa eigenbases does not induce one-loop edms.}  The phenomenological impact of this argument is discussed in Section \ref{sec:pheno}, and we summarise in Section \ref{sec:sum}. { Appendix  \ref{app:mRGE} gives the RGEs for the lepton mass matrix at dimension six in SMEFT, which are simple thanks to cancellations  resulting from the Higgs minimisation conditions. }

\section{Review}
\label{sec:rev}

{ In classical physics}, the electric dipole moment of a charge distribution  $\rho(x)$ is
a vector $\int d^3x \vec{x} \rho(x)$. For elementary particles, the dipole moments are expected to be  proportional  to   the spin   $\vec{s}$,
because it is the only available intrinsic vector;
for  a non-relativistic fermion  of spin $\vec{s}$,
 the  edm is the coefficient $d_e$, of inverse mass dimension,  
appearing in  the fermion's quantum mechanical Hamiltonian
\bea
\d H =  - d_e \frac{\vec{s}}{|\vec{s}|}\cdot \vec{E} -
d_m \frac{\vec{s}}{|\vec{s}|}\cdot \vec{B} ~~.
\label{defn}
\eea
Like the magnetic dipole moment $d_m$, it  can be observed via the EoM of the fermion, that is, an amplitude is measured  rather than a rate.

{ In the Quantum Field Theory} of a charged lepton $l$, with covariant derivative $\dslash +i eQ_l\Aslash$, the QED  interaction $eQ_l  \overline{l} \Aslash l$ implies a  magnetic  moment  of  $ \frac{eQ_l g}{2m_l}$, with $g=2$,  in the non-relativistic, quantum-mechanical Hamiltonian.
Additional small deviations  arise via QED loop corrections \cite{Schwinger:1948iu}, $(g-2) \propto \alpha/\pi+...$, where  the higher order terms may differ slightly depending on the mass of the fermion.
QED being invariant under C, P and T, it does not generate  edms,{  which are CP-odd: under the discrete  transformation CP,  $\vec{s} \to -\vec{s}$, and $ \vec{E} \to \vec{E}$.}

In the full SM, there are electroweak contributions to the electric dipole moments of the leptons,  as well as to the  magnetic dipoles. These can be parametrised below the weak scale  by  the flavour-diagonal  coefficients $\tilde{C}^{ii}_D$ of a dipole operator,  added to the Lagrangian  as
\bea
\d {\cal L}_{< m_W} &=& \sum_{ij} \frac{1}{v} \left( \tilde{C}^{ij}_{D}
 \overline{l}_i    \sigma^{\a\b}P_{R} l_j F_{\a\b}
+  \tilde{C}^{ij*}_{D}
 \overline{l_j}   \sigma^{\a\b} P_L l_i F_{\a\b} \right)~~~,\label{L}
\label{OD1}
\eea
where $\{l_j\}$ are four-component  mass eigenstate charged  leptons,  generation indices $i,j$ are summed
over $\{e,\mu,\tau\}$,   $\a,\b$ are Lorentz indices, $F_{\a\b}$ is the QED field strength,   and  $v \sim m_t$  is the vacuum expectation value of the Higgs (so $\frac{1}{v^2} = 2\sqrt{2} G_F$).
The  dipole contributions of heavy NP models  can be parametrised, at scales between $m_W\leftrightarrow \LNP$, by the  
operator  of  eqn (\ref{OD}), where $\{\ell_j\}$  and  $\{e_i\}$ are
lepton  doublets and (charged) singlets. This is the SM gauge-invariant version of the operator of eqn (\ref{OD1}), and the coefficients match at $m_W$ as
$\tilde{C}^{ij}_D = \frac{v^2}{\LNP^2}{C}^{ij}_D $. 
 Notice that the dipole operators  of eqns (\ref{OD}) and (\ref{OD1}) are defined  without a Yukawa coupling, to simplify flavour basis rotations. 

The SU(2)$\times$U(1) invariant dipole  can then be added to the SM  Lagrangian above the weak scale, along  with other dimension six  operators parametrising the NP, as 
\bea
\d {\cal L}_{SMEFT} &=& \sum_{A,\zeta} \frac{ C^{\zeta}_{A}}{\LNP^2} O_A^\zeta
+h.c.
\eea
where the operators $\{ O_A\}$ are  in the usual\footnote{In  the usual  SMEFT basis, there are $U_Y(1)$ and $SU(2)$ dipoles, whereas here the dipoles are decomposed on the $Z$ and the $\g$.}  on-shell basis \cite{BW,polonais} for SMEFT,   $\zeta$ are flavour indices,   $\LNP$ is a NP scale,  and the $+h.c.$ applies only to the operators which are not already hermitian.
The $1/\LNP^2$ factor is convenient for powercounting, but when setting experimental bounds on  coefficients, we take $\LNP = v\sim m_t$.

Recall that an operator $H$ acting on  a vector space with a hermitian inner product,
$(v_i,v_j) = (v_j,v_i)^*$,  is hermitian if and only if  $(v_i,H v_j) = (H v_j,v_i)^*$  for any $v_i,v_j$.
This definition can be applied to  operators  in Quantum Field Theory acting on  the space of states of the QFT;  in practise, for the dipole and other two-lepton operators,   it reduces to ``hermitian in flavour-space''.
So the dipole, added to the Lagrangian as $C_D O_D + C_D^\dagger O_D^\dagger$, can be decomposed into  hermitian and anti-hermitian operators (with respectively hermitian and antihermitian coefficients)  as
\bea
 C {\cal O} + C^\dagger{\cal O}^\dagger = \frac{1}{2}(C+C^\dagger)
 ({\cal O} + {\cal O}^\dagger) + \frac{1}{2}(C-C^\dagger)
 ({\cal O} - {\cal O}^\dagger) 
\nonumber
\eea
where hermitian coefficients satisfy $H = H^\dagger$  as  matrices in flavour-space.
Antihermitian operators satisfy $A = -A^\dagger$, so can be written $iH$, for $H$ hermitian.

For a lepton of flavour $j$,  the   magnetic and  electric dipole moments are proportional  to  the  flavour-diagonal hermitian and  anti-hermitian coefficient combinations(in conventions where the photon Feynman rule is $-ie Q_f \g^\m$, and the photon momentum is ingoing):
\bea
\Delta a_j = \Delta \frac{(g-2)_j}{2} &=&
{\frac{4m_j{\cal R}e[C^{jj}_D]}{e Q_jv }\frac{v^2}{\LNP^2}} ~~~ \label{Da}
\\
\Delta \frac{d}{Qe} 
&=&  - \frac{2 \Im m[C^{jj}_D]   }{ eQ_j v}  \frac{v^2}{\LNP^2}\label{de}
\eea
This distinction will be important,  because we will show that   one-loop  contributions to $[C_D]$  are of the form $[H] [m_l]$, where quantities in brackets are matrices, $[H]$ is  hermitian  in flavour space, and the charged lepton mass matrix $[m_l]$ is diagonal in the mass eigenstate  basis where edms are measured.  { As a result, there are no one-loop edms in the mass basis.}

 In the SM, 
without neutrino masses  and with $\theta_{QCD} = 0$, the edms are generated by the CKM phases.  A contribution is expected at four-loop \cite{Pospelov:2005pr,Pospelov:1991zt}, where the ``long-distance' contribution involving kaon loops  \cite{Yamaguchi:2020dsy,Yamaguchi:2020eub}
gives
\bea
{ d_e \approx 5.8 \times 10^{-40}e{\rm cm}}~~~,~~~
d_\mu \approx 1.4 \times 10^{-38}e{\rm cm}~~~,~~~
d_\tau \approx - 7.3\times 10^{-38}e{\rm cm}~~~,~~~
\label{dSM}
\eea

\renewcommand{\arraystretch}{1.3}
\begin{table}[h]
\begin{center}
\begin{tabular}{|l|l|l|}
\hline
Process & Current bound   & Future Sensitivity    \\
\hline
$BR(\mu\to e \gamma) $ & $ < 1.5 \times 10^{-13}$ \cite{MEGII:2025gzr}
		   & $  {\sim 6\times   10^{-14}}$ MEGII \cite{MEGII:2018kmf} \\
$BR(\tau\to e\gamma)$ & $<3.3\times 10^{-8}$  \cite{BaBar:2009hkt}
 & $ \sim 9 \times 10^{-9} $ BELLEII \cite{Banerjee:2022xuw} \\
$BR(\tau\to \mu\gamma)$ & $<4.2\times 10^{-8}$  \cite{Belle:2021ysv} & $\sim 6\times  10^{-9} $ BELLEII \cite{Banerjee:2022xuw} \\
$\Delta (g-2)_e/2$, & $ \lsim 8\times10^{-13}$ \cite{Morel:2020dww}
&\\
 $|d_e/(eQ)|$ & $<1.1\times10^{-29} {\rm cm}$ \cite{ACME:2018yjb} & $\sim 10^{-30} {\rm cm}$ACMEIII\cite{Ang:2023uoe} \\ 
&$<4.1\times10^{-30} {\rm cm}$ \cite{Roussy:2022cmp}&\\
$\Delta (g-2)_\mu/2$, &$\sim 160 \times 10^{-11}$ \cite{Muong-2:2025xyk}&\\
$|d_\mu/(eQ)|$ & $<1.8\times10^{-19} {\rm cm}\    \cite{Muong-2:2008ebm}$
& $\sim  10^{-21}$ cm \cite{Otani:2022wlj}\\
&& $\sim 400 \to 6 \times 10^{-23}$ cm \cite{Adelmann:2025nev} \\
$|d_\tau/(eQ)|$ & $\lesssim10^{-18}\ {\rm cm} $ \cite{Diehl:1996wm}
& $\sim  10^{-19}$ cm \cite{Belle-II:2022cgf}\\
\hline
		\end{tabular}
	\end{center} 
\caption{  Current and anticipated  experimental  constraints  on processes mediated by elements of the dipole coefficient matrix. The quoted $\Delta (g-2)_i/2$  are estimated bounds on the heavy NP contribution to magnetic moments,  obtained by subtracting the SM value \cite{Aliberti:2025beg} from the experimental determination (of $\a_{QED}$ in the electron case), and allowing  a $\sim 2\sigma$ uncertainty. 
\label{tab:bds}} 
\end{table}
 \renewcommand{\arraystretch}{1}

{ Experimental bounds} on  the  coefficients $C_D^{ij}$  arise from searches for electric and magnetic  moments of the charged leptons, as well as from upper bounds on  the  decay ratios
\bea
\frac{\Gamma(l_j\to l_i \g)}{\Gamma(l_j\to l_i \nu \bar{\nu})} &=&384\pi^2
\frac{v^6}{m_j^2\LNP^4}(|C_D^{ij}|^2 + |C_D^{ji}|^2) ~~~ \label{BR}.
\eea
(Recall that $BR(\meg) = \frac{\Gamma(\meg)}{\Gamma(\menn)}$ but the $\tau$ has hadronic decay channels, so $BR(\tlg) = \frac{\Gamma(\tlg)}{\Gamma(\tlnn)} BR(\tlnn)$, with $BR(\tlnn) \sim .176$.)

The experimental constraints listed in table \ref{tab:bds}  imply, for $\LNP=v \sim m_t$:
\bea
 C_D^{ij} \leq \left[
\begin{array}{ccc}
 1.7\times 10^{-14}  (eQ)  & 3.8\times 10^{-12}  
 &1.6\times 10^{-6}\\
  3.8\times 10^{-12}& 6.6\times 10^{-7}  (eQ)
 &2\times 10^{-6}\\
 1.6\times 10^{-6} & 2\times 10^{-6}& -
\end{array}
\right]
\eea
 where  on the diagonal is the more restrictive of the bounds on the  magnetic or electric dipoles:
\bea
{\cal I}m~ C_D^{ee} &\leq&  \frac{d_e v}{2 } 
\simeq 1.7 \times 10^{-14} (eQ_e) \\
{\cal R}e~ C_D^{ee} &\leq&  \frac{\Delta a_e v (eQ_e) }{4m_e } \simeq  6.9 \times 10^{-8} (eQ_e) \\
{\cal I}m~ C_D^{\mu \mu} &\leq&  \frac{d_\mu v}{2 }
\simeq 9.1\times 10^{-4} (eQ_\mu) \\
{\cal R}e~ C_D^{\mu \mu} &\leq&  \frac{\Delta a_\mu v (eQ_\mu) }{4m_\mu } \simeq  6.6 \times 10^{-7} (eQ_\mu) 
\eea

\section{Vanishing one-loop edms}
\label{sec:arg}

This section aims to show  that, in the models considered here, the  one-loop dipole coefficient $C_D^{ij}$  has the structure $ H^{ij} m_{l_j}$ in the lepton mass eigenstate basis.
$H$ is a hermitian matrix, so its diagonals are real meaning that edms vanish at one-loop --- although the off-diagonal  $C_D^{ij}$ which mediate $l_j\to l_i \g$ are complex.

We suppose heavy lepton-flavour-changing NP, and,
  we require  that the heavy   new particles  not  interact with the  charged  singlet leptons of the SM.
 In order to  ensure that this last  requirement is well-defined, it is implemented in the mass eigenstate basis  at the New Physics scale,  with the approximation that the electroweak scale is sufficiently small to be neglected (so all the SM particles are massless).
 This should ensure, in models with extra Higgses\footnote{CP Violation in the 2 Higgs Doublet Model is discussed in  \cite{Gunion:2005ja}, where $d_e$ has recently been studied in the decoupling limit \cite{Davila:2025goc}.} or ``leptons'' (hypercharge $-1/2$ fermionic doublets\footnote{For instance,  a fourth  generation of ``vector-like'' leptons, as discussed in \cite{Kawamura:2019rth}, or see \cite{Nortier:2024vsu,DAgnolo:2023rnh} for  phenomenological constraints.}),  that it is the light ``standard'' Higgs and leptons which participate in the Yukawa coupling with the $\{e_R^i\}$. 
This ``NP mass basis'' is  also the basis  where   matching calculations of the model onto SMEFT are performed, and has the additional advantage of being
readily accessible from  the model Lagrangian.

Section \ref{ssec:basis}  motivates our choice to  define lepton flavours as mass eigenstates, and  discusses powercounting  in loops, in  factors of $1/\LNP^2$ and in mixing angles.
Section \ref{ssec:diagrams} estimates  one-loop contributions  to the dipole in  the models considered here, and
 finds that  one-loop diagrams with  heavy new particles give vanishing edm in the mass basis, but that SM diagrams involving Yukawa couplings could be complex, because the Yukawa couplings might not be diagonal and real  in the mass eigenstate basis.
So in subsection \ref{ssec:RGEs}, we calculate the NP contribution to the lepton mass matrix at dimension six in SMEFT, which allows to calculate the Yukawas from the  known masses as a function of NP parameters, and show that the one-loop SM contributions to the edms also vanish.

\subsection{Flavour basis and powercounting}
\label{ssec:basis}


Experimentally, the flavour    of   charged leptons  is   defined  by their masses,    so in this manuscript, flavour eigenstates are  mass eigenstates.
This definition  is correct in the SM, where lepton flavours are protected by symmetry,  and it is the most relevant definition   for  flavour-changing models, because  the data is in the mass eigenstate  basis.  In addition,  it is the usual definition in EFTs below the weak scale.

The mass eigenstate definition of flavour can also be used above the weak scale.
This avoids  performing basis rotations in matching EFTs at $m_W$.
And in addition, the RGEs for the mass matrix are simpler than those for the Yukawa matrix (see Appendix \ref{app:mRGE}).

We also need Yukawas, because these are the coupling constants of Higgses with  leptons, and could appear in loop diagrams contributing to EDMs.
We calculate the Yukawa couplings as a function of the masses (and NP parameters) in the framework of SMEFT at dimension six, where the charged lepton mass matrix can be written
\beq
[m_l] = [Y_l]v - \frac{v^2}{\LNP^2}[C_{EH}]v
\label{above}
\eeq 
where the  $C^{ij}_{EH}$ are coefficients of
$O^{ij}_{EH} = H^\dagger H \overline{\ell_i} H e_j$, and  in the mass eigenstate basis,  the matrix $[m_l]$ is obviously diagonal.
The  $C^{ij}_{EH}$ can be calculated in  an NP model, and are constrained experimentally (see Appendix \ref{app:LHC}).


In the  RGEs of SMEFT, which we will use later,  Yukawas  appear  as the coupling constant of Higgs bosons in loops, but also  as  mass insertions on external legs,  because the operator basis  is reduced using the lepton  Equations of Motion (EoM).
In the case of mass insertions, we substitute the  diagonal and known mass matrix for the Yukawa $[Y_l]$,
because we are in the mass  basis.

The calculation of   dipole coefficients $C_D^{ij}$  can be performed as a perturbative expansion in  loops, scale ratios,  or other small parameters such as mixing angles.
In the loop expansion,  a contribution to the dipole  can  generically arise at one loop.
Then, in the expansion in the scale ratio $v^2/\LNP^2$,   edms and the flavour-changing dipole coefficients can first arise   at ${\cal O}(v^2/\LNP^2)$ (neglecting SM CP violation  because it is suppressed, see eqn \ref{dSM}).
However,  the diagonal  magnetic dipoles can be  unsuppressed by factors of $v^2/\LNP^2$,
because they  do not require the CP- or flavour- violating interactions that our models provide.
We therefore neglect the magnetic dipoles in this manuscript, as they can be generated by the SM and other light physics.

 The  leading contribution  to the dipole from the models  considered here could therefore  be expected  to be of order
$$
\frac{C_D^{ij}}{\LNP^2} \lsim \frac{\Pi_{ij}}{16\pi^2 \LNP^2 }\frac{m_j}{v} ~~~,
$$
where $\Pi_{ij}$ is an unknown product of NP and SM couplings, and  the lepton mass $m_j$ (or Yukawa) arises as a mass insertion on an external leg. 
But, as will be  discussed in Section \ref{ssec:diagrams}, the  NP edm  at this order vanishes.   One could then  wonder about 1-loop SM dipole diagrams, with an ${\cal O}(v^2/\LNP^2$) complex  mass correction  on an external leg, which  could be of order 
 $$
\frac{C_D^{ii}}{\LNP^2} \sim \frac{\Pi_{ii}}{(16\pi^2)^n \LNP^2 }\frac{m_i}{v} ~~~,
$$
where  $\Pi_{ii}$ is another  product of unknown couplings, and $n$ = 1 or 2, depending on whether the NP contribution to the mass matrix arises at tree or one-loop.
Notice that leptonic mass corrections are   multiplied on the right  by $[m_l]$ or $[Y_l]$, due to the peculiarity that  the $\{e_R^i\}$ only interact via Yukawas. 
So the question this manuscript aims to address, is whether   diagrams  such as those of figure \ref{fig:1loopsm} can contribute to edms, and if yes, what is their magnitude. Section \ref{ssec:RGEs} will show that  these contributions to the edm vanish in the models considered here.

\subsection{One-loop  diagrams in the  model}
\label{ssec:diagrams}

We want to  show that the charged lepton edms vanish at one loop,  provided that the new,  heavy   mass-eigenstate particles in the loop do not interact with singlet leptons.
In the converse case, where the NP does interact with singlet leptons, it is well-known that one-loop leptonic edms can arise ($eg$,  as in Supersymmetry\cite{Hisano:2008hn,Ellis:2008zy}).

We  follow  the discussion of Fujiwara etal \cite{Fujiwara:2021vam} who review and extend the Shabalin proof \cite{Shabalin:1978rs,Shabalin:1979gh}  that two-loop sunset diagrams in the SM do not contribute to  edms.
The discussion of Fujiwara etal  is in the broken electroweak   vacuum,  so corresponds to the computation of a physical $S$-matrix element.
However, we  choose to  calculate in the  SU(2)-invariant vacuum,   and use  a Lagrangian  expressed in  the mass eigenstate basis  of the heavy new particles.
This Lagrangian contains the kinetic terms, masses  and interactions of the SM and NP particles.
Therefore  the propagating particles  in the diagrams
can be identified with  fields in the  Lagrangian,  and  it is simple to
implement the restriction imposed on the NP models,
that  the new  mass eigenstates do not interact with singlet charged leptons. 
These diagrams   will contribute to top-down matching  onto the dipole operator at $\LNP$ or $m_W$.

\begin{figure}[h]
\begin{center}
\includegraphics[scale=1.2]{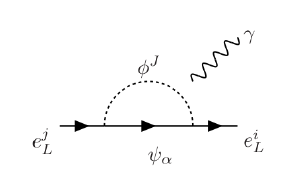}
\end{center}
\caption{One-loop diagrams that might contribute to  the dipole
operator after using EoM.  The photon attaches to either internal line.
 $\phi^J$ and $\psi_\a$ are mass eigenstate  particles of the SM or BSM in the SU(2)-invariant vaccuum.
\label{fig:1loopnp}}
\end{figure}

\begin{figure}[h]
\begin{center}
\includegraphics[scale=1.2]{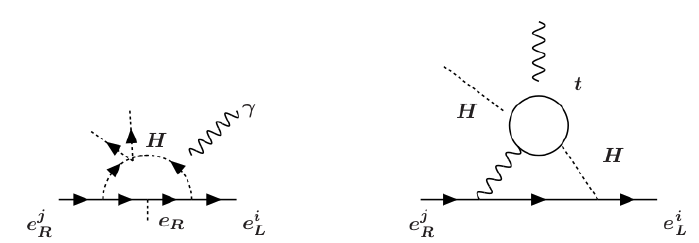}
\end{center}
\caption{Chirality-flip diagrams contributing to  the dipole
operator; since the   BSM particles do not interact with singlet charged leptons, the internal lines  of the one-loop diagram are SM particles.  As a result, this diagram matches at the electroweak scale onto the low-energy dipole of eqn (\ref{OD1}), where the extra Higgs legs do not contribute to the operator dimension.  A  two-loop ``Barr-Zee'' diagram is included  because such contributions also vanish by the same argument that suppresses the one-loop diagrams. 
\label{fig:1loopsm}}
\end{figure}

The  particles appearing in the loop are taken to be scalars $\phi_J$ or   fermions $\psi_\a$,  identified by their SU(2) representation and hypercharge. They can be  new particles or from the SM. 
 The   relevant interactions are 
\bea
\d {\cal L} &=& -y_{i \a}^J \overline{\psi}_\a \phi_J^* P_L e_i + h.c. -eQ_\a \overline{\psi}_\a \Aslash \psi_\a - ieQ_J A_\mu ( \phi_J^* \partial^\mu \phi_J -
( \partial^\mu \phi_J)^* \phi_J )
\eea
where the couplings $y_{i\a}^J$ can be complex.
The  diagrams which can contribute to the dipole are illustrated in figures \ref{fig:1loopnp} and \ref{fig:1loopsm}. 
 Since the emitted gauge boson is the photon,  the  components of SU(2) multiplets in the loop can  give different contributions, and should be calculated separately. However,   since each separate contribution  is real, as argued below, this is not a problem. 

The diagrams   with outgoing $e_{Li}$ (figure \ref{fig:1loopnp})
can contribute to the chirality-flip dipole operator via a mass insertion on an external leg. Or equivalently, in EFT,
could match  (off-shell) onto  an operator 
$\bar{\ell}^j \sigma \cdot F i \! \Dslash \ell^i$, which becomes $O_D$ after
using the  doublet EoM $ i\Dslash \ell^i - \frac{[m_l]^{ii}}{v} H e^i$.
However,  such diagrams   do not contribute to the edm,
because
$ y_{i \a}^J y_{i \a}^{J*}$   is  real for any $\a$ and $J$.
Indeed, these diagrams will  contribute $[C_D]^{ji} = [H D_m]^{ji}$ where  $[D_m]$ is the diagonal mass matrix of the charged leptons,  $H$ is a hermitian matrix $\propto \sum_{\a,J} y_{j \a}^J y_{i \a}^{J*} Q_{\a,J}$, and $Q_{\a,J}$ is  the appropriate sum of electric charges of component fields. {The scalar in Figure \ref{fig:1loopnp} could be replaced by a vector with off-diagonal interactions $g^J_{i\alpha}V^J_{\mu}(\bar{\psi}_\alpha \gamma^\mu P_L e_i)$ that could carry a complex phase. However, the same argument applies because the vector couplings $g^J_{i\alpha}$ are hermitian and the combination $g^J_{i\alpha} g^J_{\alpha i}=g^J_{i\alpha} g^{J*}_{i\alpha}$ is always real.}

In the one-loop diagram of  figure \ref{fig:1loopsm}, the scalar at the vertex with $e_R^j$ must be a Higgs, because we restrict to models where NP does not interact with $e_R$.  As a result,  the internal fermion line is a charged lepton, with a mass insertion  in order to flip the chirality three times.
In the SM,  where $[Y_l]$ is diagonal and real in the mass eigenstate basis,  this diagram gives a real contribution to the dipole coefficient (part of the electroweak contribution to $(g-2)$).

In  the presence of  NP however, the  mass eigenstate basis could be misaligned with respect to the Yukawa eigenstates  by   CP and/or flavour-changing corrections $[\d m]$ to the lepton mass matrix. In the Yukawa eigenstate basis ($[D_Y]$ is the diagonal Yukawa matrix),  the  one-loop mass matrix 
\beq
[m_l] =  [D_{Y_l}] v + [\d m] ~~~,~~~V_L^\dagger [m_l] V_R = D_m
\label{VmV}
\eeq
 can be rediagonalised with unitary transformations $V_L$ and $V_R$ on the left and right.
So the Yukawa matrix may  not be  diagonal  and real in  the mass eigenstate basis, and  the diagrams of figure \ref{fig:1loopsm} give 
\bea
\frac{v^2}{\LNP^2}C_D^{ij} &\sim& \frac{[Y_l]_{ik}m_k [Y_l]_{kj}}{16\pi^2 v}  + \frac{[Y_l]_{ij}y_t}{(16\pi^2)^2} 
\label{wrong}
\eea
where  one of the Barr-Zee two-loop diagrams is included for illustration, because these can be numerically larger than the  one-loop contribution, for heavy SM particles ($t$,$W$) in the closed loop.
Since the Yukawa matrices are not hermitian, the one-loop contribution to the edms could be non-vanishing (although suppressed by three lepton Yukawas/masses, as well as mixing angles between the mass and Yukawa eigenbases), and the Barr-Zee diagrams can contribute if the diagonal Yukawas are complex.

\subsection{Can the 1-loop SM diagrams be CP-violating?}
\label{ssec:RGEs}

We now want to show that the diagrams of figure  \ref{fig:1loopsm}  do not  contribute to  edms, because the NP contribution to the mass matrix $[\d m]$ is proportional to $[Y_l]$.
This second part of the argument is constructed in SMEFT at dimension six, which gives an explicit form for $[\d m]$.

The idea of the argument  is  to show that one-loop diagrams in the model induce mass matrix corrections of the form $[H][Y_l]$ or $[H][m_l]$ where $[H]$ is a hermitian matrix. In particular, the corrected mass matrix will be of the form $[m_l]=  [H_m] [Y_l]$, where  $[H_m]$ is hermitian. So any contribution to the dipole   of form $[C_D]= [H_D ][Y_l]$, or $[C_D]= [H_D ]([Y_l] - 3[C_{EH}]{ v^2/\Lambda_{NP}^2})$,  can be written
$
[C_D] = [H_D] [Y_l] = [H_D] [H_m]^{-1} [m_l]
$.  As a result,  in the mass eigenstate basis,   $[C_D] = [H] [D_m]$, where $H$ is hermitian, and the edms vanish.

In SMEFT at dimension  dimension $\leq 6$,  the charged lepton mass matrix is
given in eqn (\ref{above}).
The argument of the previous paragraph will apply, if $[C_{EH}] = [H][Y_l]$, where $[H]$ is a hermitian matrix.

We want to match  a NP model 
onto $O_{EH}$,  at tree level in the EFT  and at  up to  one-loop in the model
\footnote{Notice that the loop-counting  can be different  above and below the matching scale;   for instance a top and W box matches to a tree-level four-fermion operator at the weak scale. Matching at tree-level in the EFT, combined with one-loop RGEs, should give  a leading-log approximation to the full  result that is independent of the operator renormalisation scheme\cite{burashouches}.}.
Matching calculations can be  performed by equating  matrix elements  which are calculated  in the theories above and below the matching scale.
These matrix elements can involve  off-shell or on-shell particles.
For  the off-shell case,  a larger ``Green''  basis\cite{Gherardi:2020det}  of operators  should be used in the EFT,  which can then be reduced to the on-shell operator basis by applying the EoM  of the fields.
In our case of  matching  models not interacting with  charged singlets, onto an operator involving a charged singlet,   it is convenient to  first  match to the  ``Green'' basis, then  apply the EoM, because this allows to distinguish diagrams with a  dynamical Higgs in the loop, from those with a  mass insertion on an external line.

A Green basis for SMEFT, as well as the reduction of the Green coefficients  $\{G\}$ to the on-shell coefficients $\{C\}$ is given  in \cite{Gherardi:2020det}: 
\bea
[C_{EH}] &=&  -\frac{1}{2}[m_l m_l^\dagger][G_{\ell D} ][m_l]
+ [G^{'(1)}_{H\ell } ][m_l]
+ i[G^{''(1)}_{H\ell } ][m_l] +
[G^{'(3)}_{H\ell } ][m_l]
+ i[G^{''(3)}_{H\ell } ][m_l] +...
\label{GtoC}
\\
&=& [H] [m_l] +  [m_l m_l^\dagger][H] [m_l]  \nonumber
\eea
where the parameters in square  brackets are matrices in lepton flavour space,   and in the $+...$ are  numerous irrelevant  Green coefficients which are discussed further below. The mass matrices  on the right side of eqn (\ref{GtoC}) appear via the EoM. The Green  coefficients which are listed are all hermitian, and  their operators $Q$ involve a pair of doublet leptons:
\bea
Q^{ij}_{\ell D} &=& \frac{i}{2}\overline{\ell}^i \{ \Box,\Dslash\} \ell^j
\nonumber\\
Q^{'(1)ij}_{H\ell } &=& i H^\dagger H \overline{\ell}^i \!\stackrel{ \leftrightarrow}{ \Dslash} \! \ell^j 
\nonumber\\
Q^{''(1)ij}_{H\ell } &=&  \overline{\ell}^i \g_\nu \ell^j D^\nu (H^\dagger H)
\nonumber\\
Q^{'(3)ij}_{H\ell } &=& i  H^\dagger\tau^\r H \overline{\ell}^i \tau^\r
\!\stackrel{ \leftrightarrow}{ \Dslash} \!  \ell^j 
\nonumber\\
Q^{''(3)ij}_{H\ell } &=&  \overline{\ell}^i \tau^\r \g_\nu \ell^j
D^\nu (H^\dagger\tau^\r H)
\nonumber
\eea
where  $i,j$ are lepton flavour indices, $\{\tau^\r\}$ are the Pauli matrices and
$\overline{\ell}^i \!\stackrel{ \leftrightarrow}{ \Dslash} \!\ell^j =
\overline{\ell}^i  \Dslash  \ell^j
- (D_\nu \ell^i)^\dagger\g^0  \g^\nu \ell^j$.
This demonstrates that the listed contributions to  the mass correction  $[C_{EH}]$ are of the form $[H] [m_l]$, up to corrections    $[m_l m_l^\dagger][H] [m_l] $. We neglect these unwelcome terms, which  contradict the claim we aim to prove, because they are numerically small.

It remains to show that the   Green coefficients which   contribute to $C_{EH}$, but  which are not given in eqn (\ref{GtoC}),  can be neglected.
These include   bosonic   operators  involving Higgses, which can  contribute
$\propto [Y_l]$ by attaching a scalar lepton current  to a Higgs leg.  Such operators trivially give a contribution  $\propto [H] [Y_l]$.
There are also Green operators involving  $\overline{\ell^j} He^i$ and two derivatives, 
or two Higgses (the Green basis operator $Q_{EH}$). Since the only  interaction of the singlets $e^i$  is the Yukawa,  the loop diagram generating these operators will involve light particles, and therefore may contribute to the RGEs but not  in ``tree-level''  matching to the EFT.

The   contribution of SM  particle loops to $C_{EH}$
 can be included by running   the Renormalisation Group Equations from $\LNP$ to $m_W$. The RGE for $C_{EH}$ is given in eqn (\ref{appCEH}).
Eqn (\ref{appCEH})  includes terms proportional to $[C_{EH}]$,  $[Y_l]$
and  $[Y_l Y_l^\dagger Y_l]$, which are of the form $[H][Y]$.
There are also terms $[Y_l Y_l^\dagger] [C_{EH}]$ and   $[C_{EH}][ Y_l^\dagger Y_l]$,
which are neglected like the $[m_l m_l^\dagger][H] [m_l]$ terms because they are numerically small. 
The doublet lepton penguin coefficients $[C_{HL}^{(1)}]$ and  $[C_{HL}^{(3)}]$ are hermitian,  so terms of the form   $[C_{HL}^{(i)}][Y_l]$ or    $[C_{HL}^{(i)}][Y_l Y_l^\dagger Y_l]$ are acceptable. 
Since the NP does not interact with singlet charged leptons, we suppose that  neither $O_{LEDQ}$ nor $O^{(1)}_{LEQU}$ are generated in matching the model to SMEFT, so these terms can be neglected.
The remaining potentially problematic operators that mix into $[O_{EH}]$ are the singlet penguin $O_{HE} = (H^\dagger \stackrel{\leftrightarrow}{D_\nu} H) (\overline{e}\g^\nu e)$, and the four-fermion  operator $O_{LE} = (\overline{\ell}\g^\nu \ell)(\overline{e}\g^\nu e)$.
Both these operators involve vector currents of singlet charged leptons, which could  appear  in  matching  via a SM gauge current--- for instance, $O_{LE}$ could be generated by  the diagram of Figure \ref{fig:1loopnp} with the singlet current attached to the end of the gauge boson line.
However, in this case, the singlet flavour structure is the identity matrix, so $[Y_l] [C_{HE}] \propto [Y_l]$.
In $O_{LE}$, the doublet flavour structure is hermitian, as in  Figure \ref{fig:1loopnp}, so $C^{jkii}_{LE} Y_l^{ki} = [H]^{jk}[Y_l]^{ki}$.

In summary, we have shown that $[C_{EH}] = [H'][Y_l]$ (up to $m_l^2/v^2$ corrections), so $[m_l]$ of eqn (\ref{above}) is of the form $[m_l]  = [H^{''}][Y_l]v$, and the Yukawa can be written
 $[Y_l]v   = [H][m_l]$, where $H,H'$ and $H^{''}$ are hermitian.
As a result, the Higgs coupling, which is $[Y_l]$ in  the unbroken  SM, or
$[\kappa] = [Y_l] - 3 [C_{EH}]{ v^2/\Lambda_{NP}^2}$ in the true vacuum, 
can be written
\beq
\kappa^{ii}v = H^{ii} m_i \in \Re~~~,~~~ \kappa^{ik}m_k\kappa^{ki} = \frac{1}{v^2}[H D_m^2 H]D_m \in \Re
\label{lesswrong}
\eeq
where $[H D_m^2 H]$ is hermitian.
So the  diagrams estimated in eqn  (\ref{wrong})  do not contribute to the edms.

\section{Phenomenology }
\label{sec:pheno}

This section recalls  the phenomenological patterns that can arise in dipole observables,  when induced by the  models considered here.
The latter include numerous neutrino mass models, whose flavour phenomenology \cite{Chun:2003ej,Kakizaki:2003jk,Akeroyd:2009nu,Dinh:2012bp,Barrie:2022ake,Tommasini:1995ii,Ilakovac:1994kj,Dinh:2012bp,Alonso:2012ji,Abada:2015oba,Coy:2018bxr,Abada:2021zcm,Granelli:2022eru,Crivellin:2022cve,Ibarra:2011xn,Petcov:2005jh,Toma:2013zsa,Ardu:2023yyw,Ardu:2024bua,CentellesChulia:2024iom,Batra:2023mds,Zhou:2021vnf,Giarnetti:2023osf,Abada:2008ea,Forero:2011pc,Hagedorn:2018spx}, and edm predictions \cite{Ng:1995cs,Archambault:2004td,Abe:2013qla,Abada:2018zra,Abada:2024hpb,Fujiwara:2021vam,Fujiwara:2020unw} have been widely studied.
The patterns are identified by expanding the dipole coefficients in loops, in $v^2/\LNP^2$ and in the charged lepton masses.
This means that the patterns  are expected, but not obligatory, because they could fail due to hierarchies or symmetries in the NP couplings, or due  to  accidental cancellations.

The previous Section showed that  contributions to edms can first  arise at two-loop, and are proportional to the relevant lepton mass,  as parametrised in Eq.~(\ref{estimate}). The radiative decays $l_j \to l_i \g$ can be mediated by the one-loop dipole  coefficient matrix, which  at dimension six  reads
\begin{equation}\label{Dipolematrix}
	C_{D}= \frac{1}{16 \pi^2 \,v} [H][D_m] =  \frac{1}{16 \pi^2 \,v} \begin{pmatrix}
		H_{ee} m_e & H_{e\m} m_\mu & H_{e\tau} m_\tau \\
		H_{e \mu}^{*} m_e & H_{\mu \mu} m_\mu & H_{\mu \tau} m_\tau \\
		H_{e \tau}^{*} m_e & H_{\mu \tau}^{*} m_\mu & H_{\tau \tau} m_\tau \\
	\end{pmatrix} ~~~,
\end{equation}
where $H$ is a  hermitian matrix  depending on NP and possibly SM couplings.

The  edms follow the  scaling $d_{i}\propto m_{i} \Im m\{\Pi_{ii}\}$,
where $\Pi_{ii}$ is a product of NP and SM couplings. If $\Im m\{\Pi_{ee}\}$ is comparable to  $\Im m\{\Pi_{\mu\m}\}$ and   $\Im m\{\Pi_{\tau\tau}\}$,
then the stringent electron edm bound shown in Table~\ref{tab:bds} implies that $d_{\mu}$ and $ d_{\tau}$  are well below the current and future experimental reach.
In addition, the edms are suppressed by two loops.  As a result, for  $\Lambda_{NP} \gtrsim $ 1 TeV,  $d_{\mu}$ and $d_{\tau}$ cannot exceed values around $10^{-25}\, {\rm e\cdot cm}$ and $10^{-24}\, {\rm e\cdot cm}$, which is below the sensitivity of future searches.  So an observation of   $d_{\mu}$ or $d_{\tau}$ would  be a footprint of models with more assertive CP violation. 

The electron edm is known to probe higher 
NP scales

\begin{figure}[t]
	\begin{center}
		\includegraphics[scale=0.5]{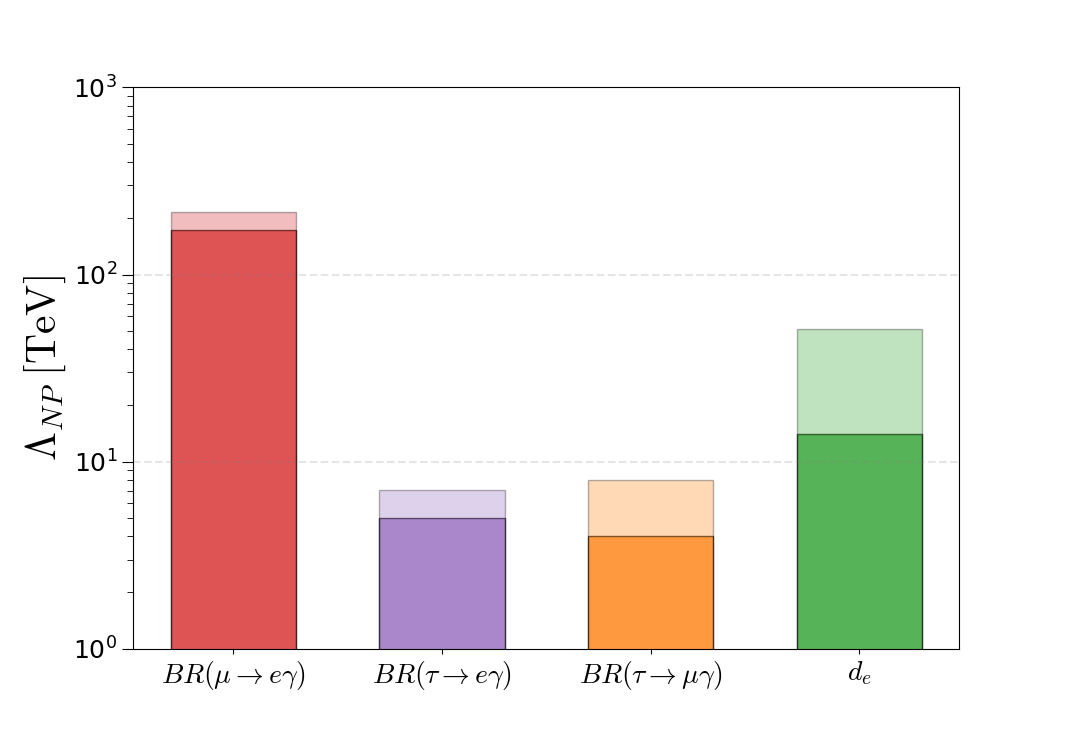}
	\end{center}
	\caption{The NP scale probed by the observables on the horizontal axis,
          in lepton flavour-changing models where the NP does not interact with singlet charged leptons. The darker bars are obtained from  the current experimental bounds  in Tab.\ref{tab:bds} (future projections are shaded),  assuming the coupling constant combinations  $H_{ij}$ and  $\Im m\{\Pi_{ii}\}$ are ${\cal O}(1)$. 
		\label{fig:barplot}}
\end{figure}

Sthan almost any other observable\cite{EuropeanStrategyforParticlePhysicsPreparatoryGroup:2019qin}, where this reach is obtained  by assuming ${\cal O}(1)$ CP-violating couplings, and no suppression by loop factors or Yukawa couplings.
However, in the class of models considered here,  the NP scale  probed by  $d_e$ could be lower than that of  $\meg$, as illustrated in  figure~\ref{fig:barplot}.
This  relative suppression arises due to the lepton mass factors combined with the  extra loop suppression of  edms :
\begin{equation}
  \Im m	\{C^{e e}_D\}  \sim \frac{1}{16\pi^2}\frac{m_{e}}{m_\m} \,
  \left(\frac{  \Im m \{\Pi_{ee}\} }{H_{e\m}}\right)   C^{e\mu}_D ~~~, 
\end{equation}
where the quantity in parentheses is unknown.

There are also patterns  expected among the  decays $l_j\to l_i \g$,  due to the mass factors appearing in Eq. (\ref{Dipolematrix}). Firstly, the three   branching ratios for $\meg$, $\tmg$ and $\teg$  can be comparable, because the rate ratio of Eq (\ref{BR}),
\beq
\frac{\Gamma(l_j\to l_i \g)}{\Gamma(l_j\to l_i \nu \bar{\nu})}
\simeq 384 \pi^2 \frac{v^4}{\LNP^2}(|H_{ij}|^2 + \frac{m_i^2}{m_j^2}|H_{ji}|^2)
\label{BRpheno}
\eeq
is controlled by the  coupling constant combination appearing in $H_{ij}$. 
If  the off-diagonal elements of $[H]$ are comparable, then the current bound on $\meg$ from MEGII \cite{MEGII:2025gzr}  could preclude observations of $\tau \to l \gamma$ at Belle II. This is illustrated in Figure~\ref{fig:barplot}.

The second expected pattern in the radiative decays concerns the polarisation of the outgoing lepton, which can be measured from the angular distribution \cite{Kuno:1999jp,Kitano:2000fg} when the decaying lepton is polarised (as is the case for  MEGII and  BelleII).
The terms in  the parentheses of Eq. (\ref{BRpheno}) correspond  respectively to outgoing ``left-handed'' and ``right-handed'' leptons(see Eq. \ref{BR}), and since
$|H_{ij}|=|H_{ji}|$ because  $[H]$  is hermitian, the rate to right-handed leptons is expected to be suppressed by $m_i^2/m_j^2$.

To conclude  this section, recall that these phenomenological expectations neglect potential cancellations among NP parameters.  Symmetries can be  inposed to inhibit  flavour change among the first two generations \cite{Barbieri:2011ci}, which could suppress $\meg$  while allowing  an observable $d_e$ for  $\vert H_{e \m}\vert /{\Im }m\{\Pi_{ee}\} \lesssim 0.03$. Similarly,  an approximate  CP symmetry could  suppress the electron edm. Also, accidental cancellations can occur: in the Type II Seesaw,  References \cite{Ardu:2023yyw,Ardu:2024bua}  showed that accidental cancellations between one and two-loop contributions to $\mu_R\to e_L \g$ could suppress the rate below the reach of MEGII,  for model  parameters allowing $\tau \to l \g$ to be observed at Belle II.

\section{Summary} 
\label{sec:sum}
Despite that electric dipole moments (edms) and  lepton flavour change  probe the violation of different symmetries--- meaning that they can be independently suppressed in models---they meet up in the dipole operator of eqn~(\ref{OD}), where they appear to be distinguished merely by their flavour indices. 
In CP and flavour-changing models, one naturally expects the model to generate  corrections to the lepton mass matrix, as well as the coefficient matrix of the dipole operator.
Both the mass matrix and the dipole are chirality-flip and non-hermitian, meaning they are transformed independently on the left and right under flavour basis redefinitions.
So  one could expect that  rediagonalising the mass matrix (see eqn~(\ref{VmV})),  could  mix  the $(l_j\to l_i\g)$ amplitudes with the edms, with potentially curious phenomenological consequences.
However, as shown in this paper, this naive expectation fails  in the models  considered here.

We focus on CP and lepton flavour-changing heavy  New Physics (NP) models, where the new particles interact with  lepton doublets (so as to generate neutrino masses),   but  not    with the SM  singlets $\{e_R^i\}$. In such models, the one-loop diagrams contributing to edms are illustrated in figures  \ref{fig:1loopnp} and  \ref{fig:1loopsm}, and 
Section \ref{sec:arg} argues that the dipole coefficient matrix is of the form $[C_D]\sim [H] [m_l]$, where $H$ is a hermitian matrix, so the one-loop edm vanishes in  both the  mass and Yukawa  eigenstate bases.

The argument proceeds in two steps:  Section \ref{ssec:diagrams} shows that  one-loop matching contributions at the scale of heavy new physics  follow the $[H][m_l]$ structure, since the relevant combination of couplings is hermitian and the chirality flip originates from an external mass insertion.
Then Section \ref{ssec:RGEs} shows that   NP-induced mass corrections---which  could misalign the Yukawa and mass bases--- share the $[H] [m_l]$ structure. 
So also  diagrams where  the chirality flip is via   Higgs  couplings,  which are potentially complex or off-diagonal in the mass basis,  give vanishing contributions to the  one-loop  edms, as shown in eqn~(\ref{lesswrong}).

Finally, Section~\ref{sec:pheno} recalls  the phenomenological expectations for  dipole observables  in this class of models, which are summarized in figure \ref{fig:barplot}.

\appendix
\section{The SMEFT RGEs for the charged lepton  mass matrix up to dimension six} \label{app:mRGE}

This Appendix gives  the  RGEs for  Yukawa matrix $[Y_l]$   and the dimension six ``Yukawa correction''  matrix $C_{EH}$, and shows that
the resulting  RGEs for the mass matrix $[m_l] = [Y_l]v -\frac{v^2}{\LNP^2} [C_{EH}]v$  are  simpler.

If the SM Higgs potential is written $\d {\cal L} =  + \m_H^2 H^\dagger H -\frac{\lambda}{2} (H^\dagger H)^2$,
then the RGEs for $Y_l$ at 1-loop in SMEFT above $m_W$  are :
\bea
16\pi^2 \mu \frac{\partial}{\partial \mu}[Y_l]&=&
\frac{3}{2}[Y_l Y_l^\dagger Y_l]
+ \left[Tr\left \{ 3 [Y_uY_u^\dagger] + 3[Y_dY_d^\dagger] + [Y_lY_l^\dagger]\right\}
-\frac{9(g_1^2 +g_2^2)}{4} \right]
[Y_l] \nonumber \\
&&+\frac{2\mu_H^2}{\LNP^2}\left( 3[C_{EH}] -(C_{H\Box} -\frac{1}{2} C_{HD})[Y_l]
+ [(C_{HL}^1 + C_{HL}^3)Y_l]- [Y_l C_{HE}] \right. \nonumber\\
&&~~~~~~~\left.
-2C_{LE}^{ab jk} Y_l^{bj} + N_c C_{LEDQ}^{abpq}Y_d^{qp} -N_cC_{LEQU}^{abpq} Y_u^{pq *} \right)
\label{appY}
\eea
where $\mu$ is the renormalisation scale, the first two terms are the  SM part \cite{Machacek:1983fi,Luo:2002ey}, and the contribution from dimension six SMEFT operators \cite{JMT1,JMT2,JMT3} is  in the Warsaw basis\cite{polonais,BW} and neglects gauge interactions.

The RGEs for  $C_{EH}$ are \cite{JMT1}:
\bea
16\pi^2 \mu \frac{\partial}{\partial \mu}[C_{EH}]&=&2 \lambda\left( 3 [C_{EH}]
- (C_{H\Box} -\frac{1}{2}C_{HD}) [Y_l]
+  ([(C^1_{HL} + C^3_{HL}) [Y_l]  -[Y_l C_{HE}]) \right.\nonumber\\
&&\left.-2C_{LE}^{ab jk} Y_l^{bj} + N_c C_{LEDQ}^{abpq}Y_d^{qp} -N_cC_{LEQU}^{abpq} Y_u^{pq *} \right)  + 6 \lambda [C_{EH}] + 4 \lambda  {[C^3_{HL}][Y_l]} \label{appCEH}\\
&&+~~  2 \eta  [Y_l] +  [Y_l Y_l^\dagger Y_l](C_{HD} -6C_{H\Box})
+2[C_{HL}^{(1)}] [Y_l Y_l^\dagger Y_l] -2 [Y_l Y_l^\dagger Y_l][C_{HE}]
\nonumber\\
&&+8 C_{LE}^{\a\r\s\b} [Y_l Y_l^\dagger Y_l]^{\r\s} -4N_c [C_{LEDQ}] [Y_{d} Y_{d}^\dagger Y_{d}] +4N_c [C_{LEQU}^{(1)}] [ Y_{u}^\dagger Y_{u} Y_{u}^\dagger]\\
&&{+4[C_{EH}][Y^\dagger_l Y_l]+5[Y_l Y_l^\dagger][C_{EH}]}
\nonumber
\eea
where  gauge interactions and dimension eight contributions $\propto  \mu_H^2/\LNP^4$ are neglected,
 and $\eta =  Tr\{[C_{EH} Y_l^\dagger]\} -2Tr\{ [C_{HL}^{(3)}]Y_lY_l^\dagger]\}$ (This is $\eta_1 + \eta_2 + i\eta_5$ of eqn A.3 of \cite{JMT2}, neglecting the quark operators).

The  RGE for the mass matrix can be obtained  by subtracting
$\frac{v^2}{\LNP^2} \times $eqn (\ref{appCEH}) from eqn(\ref{appY}).
One observes that there is a cancellation between the terms in parentheses of eqns (\ref{appCEH}) and (\ref{appY}), after  using the Higgs potential minimisation condition $\mu_H^2 = \lambda v^2$ (for $v = \langle H \rangle$). This cancellation arises because  the diagrams  corresponding to the operators in parentheses give  ``Green'' operators involving $\Box H$, which reduces to the SMEFT operators in the pair of parentheses after using the Higgs EoM
$$
D_\a D^\a H  + \mu_H^2 H -  \lambda H^\dagger H H = 0
$$
Since $\Box H$ vanishes in the broken electroweak vacuum,  it is normal that these diagrams do not contribute to running the charged  lepton mass matrix.  As a result, the RGE for the mass matrix is 
\bea
16\pi^2 \mu \frac{\partial}{\partial \mu} [Y_l- \frac{v^2}{\LNP^2} C_{EH}]
&=& \frac{3}{2}[Y_l Y_l^\dagger Y_l]
+ {\Big[}Tr\{ 3 [Y_uY_u^\dagger] + 3[Y_dY_d^\dagger] + [Y_lY_l^\dagger]\}
-\frac{9(g_1^2 +g_2^2)}{4} {\Big]} [Y_l] \nonumber\\
&& -\frac{v^2}{\LNP^2} \left[ 2 \lambda {\Big (} 3  [C_{EH}] +2[C^3_{HL}][Y_l] {\Big )} +4[C_{EH}][Y^\dagger_l Y_l]+5[Y_l Y_l^\dagger][C_{EH}] \right.  \nonumber\\
&&+~~  2 \eta  [Y_l] +  [Y_l Y_l^\dagger Y_l](C_{HD} -6C_{H\Box})
+2[C_{HL}^{(1)}] [Y_l Y_l^\dagger Y_l] -2 [Y_l Y_l^\dagger Y_l][C_{HE}]
\nonumber\\
&&+\left. 8 C_{LE}^{\a\r\s\b} [Y_l Y_l^\dagger Y_l]^{\r\s} -4N_c [C_{LEDQ}] [Y_{d} Y_{d}^\dagger Y_{d}] +4N_c [C_{LEQU}^{(1)}] [ Y_{u}^\dagger Y_{u} Y_{u}^\dagger] \right]
\eea
So one sees that the dominant running of the mass matrix is  probably due to the SM gauge and top couplings $\propto (3y_t^2 -9(g_1^2 + g_2^2)/4) [Y_l]$, modulo the NP term $\propto  \frac{v^2}{\LNP^2} 12 y_t^3C_{LEQU}^{(1)ijtt}$.

\section{LHC constraints on $C_{EH}^{ij}$}
\label{app:LHC}

The LHC experiments  observe $h \to \tau\bar{\tau}$ and $h\to \mu\bar{\mu}$ \cite{CMS:2020xwi,ATLAS:2025coj},
and   search for $h\to e\bar{e}$ \cite{CMS:2022urr}
and flavour-changing Higgs decays $h\to l_i^\pm l_j^\mp$ \cite{CMS:2021rsq,ATLAS:2023mvd,CMS:2023pte}.
This constrains the magnitudes of the $C_{EH}$ coefficients:
\bea
[|C_{EH}|] \lsim  \left[\begin{array}{ccc} 
2\times 10^{-4} &  8.5\times 10^{-5} &  6\times 10^{-4}\\
8.5\times 10^{-5} & 3\times 10^{-5}  &  4.9\times 10^{-4}\\
 6\times 10^{-4}&  4.9\times 10^{-4}& 2.5\times 10^{-4}
\end{array}\right]
\sim
 \left[\begin{array}{ccc} 
74 \frac{m_e}{v} &  0.12 \frac{m_\mu}{v} &  0.06\frac{m_\tau}{v}\\
 24\frac{m_e}{v} &0.05\frac{m_\mu}{v} &  0.05\frac{m_\tau}{v}\\
  204\frac{m_e}{v}&  0.83\frac{m_\mu}{v}&0.025\frac{m_\tau}{v}
\end{array}\right] \nonumber ~~~.
\eea

\end{document}